\documentclass[floatfix,twocolumn,preprintnumbers,nofootinbib]{revtex4}
\pdfoutput=1
\usepackage{graphicx}
\usepackage{dcolumn}
\usepackage{bm}
\usepackage{amsfonts}
\usepackage{amsthm}
\usepackage{amsmath}
\usepackage{amssymb}

\newcommand{\bea}{\begin{eqnarray}}
\newcommand{\eea}{\end{eqnarray}}
\newcommand{\be}{\begin{equation}}
\newcommand{\ee}{\end{equation}}
\newcommand{\bi}{\begin{itemize}}
\newcommand{\ei}{\end{itemize}}

\newcommand{\beq}{\begin{equation}}
\newcommand{\eeq}{\end{equation}}

\begin{document}

\title{Is the PAMELA Positron Excess Winos?}

\author{Phill Grajek}
\email{phillip.grajek@umich.edu}
\affiliation{Michigan Center for Theoretical Physics
University of Michigan, Ann Arbor, Michigan 48109, USA.}
\author{Gordon L. Kane}
\email{gkane@umich.edu}
\affiliation{Michigan Center for Theoretical Physics
University of Michigan, Ann Arbor, Michigan 48109, USA.}
\author{Dan Phalen}
\email{phalendj@gmail.com}
\affiliation{Michigan Center for Theoretical Physics
University of Michigan, Ann Arbor, Michigan 48109, USA.}
\author{Aaron Pierce}
\email{atpierce@umich.edu}
\affiliation{Michigan Center for Theoretical Physics
University of Michigan, Ann Arbor, Michigan 48109, USA.}
\author{Scott Watson}
\email{watsongs@umich.edu}
\affiliation{Michigan Center for Theoretical Physics
University of Michigan, Ann Arbor, Michigan 48109, USA.}

\date{\today}

\begin{abstract}
{
Recently the PAMELA satellite-based experiment reported an excess of galactic positrons that could be a signal of annihilating dark matter. The PAMELA data may admit an interpretation as a signal from a wino-like LSP of mass about 200 GeV, normalized to the local relic density, and annihilating mainly into W-bosons.  This possibility requires the current conventional estimate for the energy loss rate of positrons be too large by  roughly a factor of five.  Data from anti-protons and gamma rays also provide tension with this interpretation, but there are significant astrophysical uncertainties associated with their propagation. It is not unreasonable to  take this well-motivated candidate seriously, at present,  in part because it can be tested in several ways soon.  The forthcoming PAMELA data on higher energy positrons and the FGST (formerly GLAST) data, should provide important clues as to whether this scenario is correct.  If correct, the wino interpretation implies a cosmological history in which the dark matter does not originate in thermal equilibrium.
}
\end{abstract}
\maketitle

\section{Introduction}
Recently, the PAMELA collaboration (a Payload for Anti-Matter Exploration and Light-nuclei Astrophysics) released preliminary results \cite{Adriani:2008zr} indicating an excess of cosmic ray positrons above the $10$ GeV energy range.
This confirms earlier results from HEAT \cite{Barwick:1995gv,Barwick:1997ig} and AMS-01 \cite{AMS01}, which had already received some initial interest from theorists, e.g.,  \cite{Baltz:2002ua,Baltz:2001ir,Kane:2001fz}.   

One possible explanation for the positron excess is the annihilation of weakly interacting massive particles (WIMPS) in the galaxy.  A spate of new particle physics models have been proposed, in part to fit the detailed features of the PAMELA data \cite{NewModels,list2}.  However, it is worth exploring in detail whether a well-established candidate (such as the neutralino) could plausibly fit the data.  As discussed recently in \cite{Grajek:2008jb},  for this explanation to be valid the neutralinos would need to have a larger cross section than dark matter of thermal origin.

In light of the preliminary findings of PAMELA, we revisit the non-thermal neutralino models considered in \cite{Grajek:2008jb,Nagai:2008se,Moroi:1999zb,Kane:2001fz} to determine whether they could account for the excess\footnote{See also a recent discussion in \cite{Ishiwata:2008cv}.}.  We find a wino-like neutralino with mass roughly $200$ GeV comes close to accounting for the excess, but only if  unconventional assumptions about the underlying distribution of the dark matter or the propagation of its annihilation products are made.  Without such modifications, a light supersymmetric particle appears unable to account for the data. In part, the purpose of this paper is to point out the places where the sources of tension arise for this explanation, while simultaneously highlighting the types of astrophysical modifications that would need to be made to accomodate the data.  Such a candidate is well motivated theoretically.  For example, a wino LSP arises in theories where the anomaly--mediated\cite{AMSB} contribution to the gaugino masses dominates, including simple realizations of split supersymmetry, and the string constructions where $M$-theory is compactified on a $G_{2}$ manifold\cite{G2}.

A pure-wino 200 GeV neutralino annihilates dominantly to $W$-bosons, with a cross section $\langle \sigma v \rangle = 2  \times$ 10$^{-24}$ cm$^{3}$ s$^{-1}$.  It is remarkable that this cross section is approximately the correct one needed to explain the size of the signal in the data. Masses somewhat below 200 GeV could conceivably explain the spectrum from the positron data, but such candidates come into sharper conflict with the existing limits from anti-protons and gamma rays. Even at 200 GeV, a wino has tension with the existing data, a fact implicit in \cite{Cirelli:2008pk,Donato:2008jk}. In fact, taking the data at face value, such a candidate is excluded.  In the following, we will show how close the 200 GeV neutralino comes to the current data, given the present understanding of the astrophysics.  Given the inherent astrophysical uncertainties, it is not unreasonable to think the $200$ GeV case might ultimately be consistent with existing positron, anti-proton and $\gamma$-ray data.    Neutralino masses much larger than this
give a bad fit to the PAMELA results unless very large astrophysical boost factors are employed \cite{ZurekHooper}.  

\section{Positron Excess from Neutralino dark matter  \label{section1}}

Dark matter annihilations produce a differential flux of cosmic rays  per unit time, energy and volume as 
\be \label{source}
Q(E,\vec{x}) = \frac{1}{2} \langle \sigma_xv \rangle \left(  \frac{ \rho_x(\vec{x}) }{m_x} \right)^2 \sum_i B_i \, \frac{dN^i}{dE},
\ee
where $E$ is kinetic energy of the cosmic rays, $\langle \sigma_xv \rangle$ is the thermally averaged annihilation cross section and velocity of the non-relativistic dark matter, $B_i$ and $dN^i/dE$ are the branching ratios and fragmentation functions, and the sum is taken over all annihilation channels which contain positrons in the final state.  The observed flux of cosmic rays is then found by propagating the source of Eq.~(\ref{source}), along with 
any astrophysical sources of cosmic rays (background) to the Earth.
There are three sources of uncertainty for the prediction of any Dark Matter signal: the dark matter distribution, the propagation of its annihilation products,  and the role of astrophysical backgrounds.

Unless otherwise stated,  we restrict our attention to the commonly adopted Navarro-Frenk-White (NFW) profile \cite{Navarro:1996gj}.  This distribution is spherically symmetric and has the form:
\begin{equation} \label{nfw}
\rho(r) = \rho_{\odot} \left(\frac{r_{\odot}}{r} \right) \left(\frac{1+\left(r_{\odot}/r_s \right)}{1+\left(r/r_s\right)} \right)^2,
\end{equation} 
with $r_s = 20$ kpc, where $r_{\odot} = 8.5$ kpc is the galactocentric distance of the sun and $\rho_{\odot} = 0.3$ GeV/cm$^3$ is the local dark matter density.   
Any reasonable profile will not effect the positrons appreciably, since they are mostly local.  However, constraints from EGRET (and predictions for FGST) are directly effected by this choice, as well as fluxes from anti-protons.  While the choice of profile does not appreciably effect the positron spectrum,
the local sub-structure (clumpiness) of the dark matter,  could have important effects, as we will soon discuss.  

For propagation of the cosmic rays resulting from WIMP annihilations we use DARKSUSY  \cite{DarkSUSY} and GALPROP \cite{Strong:1998pw}, to numerically solve for the fluxes.  We vary the propagation parameters to examine how well a light neutralino can account for the positron excess. 
As standard values we take a diffusion coefficient of $K=3 \times 10^{27} \epsilon^{0.6}$ cm$^2$ s$^{-1}$, a half height for the confinement region of $L=4$ kpc, and an energy loss time of $\tau=10^{16}$ s. For the astrophysical background in positrons we adopt the power-law $\Phi = \left( 4.5  \epsilon^{0.7} \right) / \left(  1+650 \epsilon^{2.3} + 1500 \epsilon^{4.2} \right)$ from \cite{Baltz:1998xv}, where $\epsilon$ is the energy in units of GeV.

\begin{figure}
\includegraphics[width=\columnwidth]{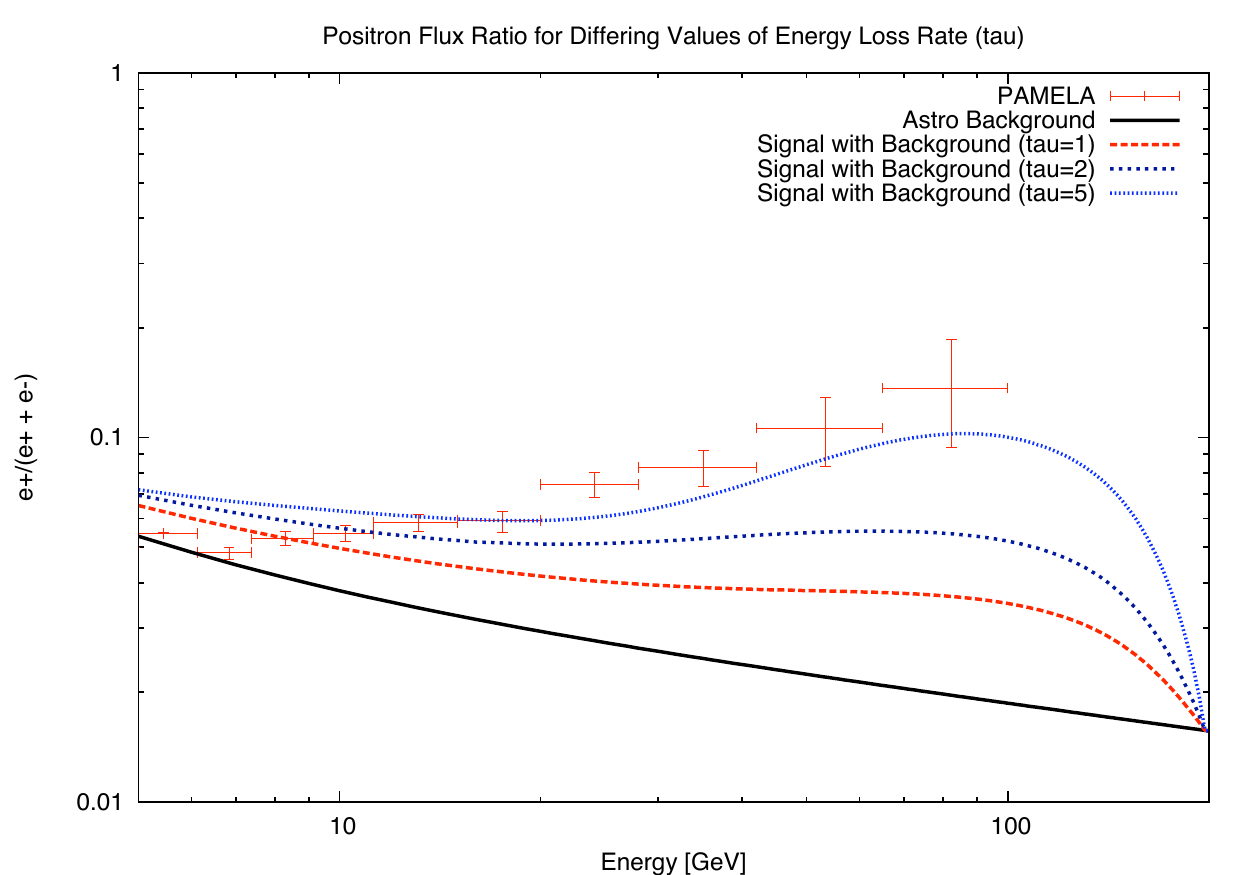}
\caption{\label{figurepos} Positron flux ratio for Wino-like Neutralino with a mass of $200$ GeV, normalized to the local relic density.
We set the height of the propagation region at $4$ kpc and consider varying values for the energy loss rate ($\tau=1,2,5$) in units of $10^{16}$ s. The solid bottom line represents a conventional astrophysical background \cite{Baltz:1998xv}.}
\end{figure}

For comparing theoretical predictions for positrons with the data, it is customary to consider the positron fraction, which includes both contributions from dark matter $\Phi^{DM}_{e^+}$ as well as the astrophysical background $\Phi^{bkg}_{e^+}$ and the analogous fluxes for electrons, i.e.
\be
\Phi=\frac{ \Phi_{e^+}^{DM}+\Phi_{e^+}^{bkg} } {\Phi^{DM}_{e^+}+\Phi^{bkg}_{e^+}+\Phi^{DM}_{e^-}+\Phi^{bkg}_{e^-}},
\ee
  
Our results for a $200$ GeV particle that annihilates to $W$ bosons  appear in Figure \ref{figurepos}. 
We have set the annihilation cross section, by assuming that the particle is a pure wino.  
The preliminary data of  PAMELA are also shown \cite{Adriani:2008zr}, along side the anticipated signal in the presence of a 200 GeV wino\footnote{  
Below $10$ GeV, the effects of charge bias on the solar modulation are expected to be non-negligible.  We have checked that solar modulation \cite{clem} can bring the PAMELA data into improved agreement with the theoretical estimate, but do not attempt a detailed accounting of the solar modulation which would require additional detailed data on the solar B field.  As the dark matter signal is dominant at higher energies, we focus on the region above 10 GeV where these effects are not important.}.   The bottom (solid) curve represents the astrophysical background.  The next higher curve represents the $200$ GeV wino for the NFW halo profile and default propagation parameters discussed above.  The dark matter signal does not provide a convincing explanation for the excess reported by PAMELA.

However, we find that by varying the rate of energy loss of the positrons a better fit to the data is possible.
Positrons lose energy via synchrotron radiaton and via inverse compoton scattering off diffuse starlight and the cosmic microwave background.  These energy losses are parameterized by the energy loss time $\tau$.  At the energies of interest this is dominated by the interaction of the positrons with starlight.   While have seen that the typically chosen default value of  $\tau =10^{16}$ s gives a poor fit to the data,  claims in the literature \cite{Olzem:2007zz}, indicate that there are theoretical uncertainties in $\tau$ at the level of a factor of 2.  So, we provide a curve for $\tau = 2 \times 10^{16}$ s, still a poor fit.  A $\tau= 5 \times 10^{16}$ s gives a qualitatively good fit to the data. It is unclear that such a value is consistent with extant maps of starlight\cite{moskalenko}. To clarify whether a neutralino could fit the data, it is important to determine this with certainty.  

The distribution of the Dark Matter could also significantly impact the fit.  Depending on the distribution of the dark matter (e.g., if there are significant over densities of the Dark Matter close by) astrophysical boost factors could also contain a dependence on the energy (see e.g. \cite{Lavalle:1900wn}).  This would largely mimic the effects of a change in $\tau$, and could act to change the spectrum from the dark matter annihilation.  An extreme example of this effect appeared recently in \cite{ZurekHooper}, where a local clump of 800 GeV wino Dark Matter was able to give the desired spectrum.  We stress that the results appearing in Figure \ref{figurepos} do not include any astrophysical boost factors. 

In summary, allowing for uncertainties in the energy loss rate and/or allowing for a small energy dependent boost factor may lead to an {\em effective} value of $\tau$ that could allow the $200$ GeV candidate to account for the excess reported by PAMELA.   Without invoking these uncertainties,  an additional source of positrons is required.

\section{Existing Constraints \label{section2}}
Strong bounds are set by existing data from $\gamma$-ray and anti-proton measurements. However both suffer from a number of uncertainties.
\subsection{Gamma Ray Constraints}
We begin with a brief review of $\gamma$-ray fluxes coming from dark matter annihilations, which are sensitive to both the 
halo profile and the diffuse $\gamma$-ray background.         
We then discuss existing constraints coming from the Energetic Gamma Ray Experiment Telescope (EGRET), which observed $\gamma$-rays coming from the galactic center.

\subsubsection{Overview of $\gamma$-rays from dark matter Annihilation}
We are interested in the continuum energy spectrum of gamma-ray flux coming from neutralino annihilations.
The differential flux is given by
\bea  \label{diffflux}
\frac{d^2\Phi_\gamma}{d \Omega dE_{\gamma}} & = & \frac{\langle \sigma v\rangle }{8\pi m_\chi^2} \sum_f \frac{dN_f}{dE_{\gamma}} B_f \int_{l.o.s}  \rho^2(l) dl(\psi), \;\;\; \;\;\; \label{eq:flux}
\eea 
which is in units of photons$/$cm$^{2}$$/$s$/$GeV$/$steradian (sr).

The first factor
depends only on the particle physics. $\langle \sigma v \rangle$ is the thermally averaged product
of the annihilation cross section. $dN_f/dE_{\gamma}$ is
the differential photon yield for a particular decay with branching ratio $B_f$, and the sum
is taken over all relevant decays.  The second piece contains the distribution of dark
matter, where $\rho(l)$ is the dark matter halo density profile and the integral is performed
along the line of sight that originates from our location in the galaxy and continues through
the full extent of the halo at an angle $\psi$ relative to the ray passing through the galactic-center.

To isolate astrophysical uncertainties it is convenient to introduce the dimensionless function $J(\psi)$
\begin{eqnarray}
\label{J}
J(\psi) & \equiv & \frac{1}{r_\odot \rho_\odot^2}  \int_{l.o.s}  \rho^2(l) dl(\psi).
\end{eqnarray}
Ground and satellite based detectors will observe a finite patch of the sky with a given angular resolution.
Therefore, when comparing theoretical predictions with what may be detected, we should average $J$ over the minimum angular resolution of the detector,
\be \label{javg}
\langle J \rangle  =  \frac{1}{\Delta\Omega}\int J(\psi)d\Omega
\ee
where $\Delta \Omega$ is the angular resolution (in steradians). This value is dictated by the experiment, e.g.  this corresponds to $\Delta \Omega = 10^{-3}$ sr for EGRET, and $\Delta \Omega =  10^{-5}$ sr for FGST. 
Given the minimum angular resolution, the dark matter profile, and the source location we can perform the average in (\ref{javg}) using e.g. DarkSUSY. Some results for the line of sight integral to the galactic center appear in Table 1.
An examination of Table 1 shows that the difference between a flat profile (Isothermal Cored) and NFW for EGRET can introduce two orders of magnitude difference in the signal.   We also show the $\langle J \rangle$ for the Einasto profile, which has recently been favored by N-body simulations \cite{Navarro:2008kc}.

While isothermal cores are now disfavored by N-body simulations, it is still fair to say that the current lack of knowledge of the halo profile induces a large error in the predicted flux from the galactic center. 

\begin{table}
\begin{tabular}{|c | c| c|}
\hline
Profile & EGRET \& &  FGST\\
&   Ground Based&  ($\Delta \Omega=10^{-5}$ sr)  \\
&  ($\Delta \Omega=10^{-3}$ sr) &\\
\hline
Isotherm & $30$ & $30$ \\
\hline
NFW  &  $1,214$ & $12,644$  \\
\hline
Einasto & $ 1,561 $ & $5,607$ \\
\hline
\end{tabular}
\caption{The averaged line of site integral $\langle J \rangle$ to the galactic center for the NFW, Einasto, and Isothermal profiles with EGRET and FGST minimal resolution.}  

\end{table}

Using the expression for the flux (\ref{eq:flux}) and averaging over the angular acceptance, the differential flux measured in the detector is
\bea
\frac{d \Phi_\gamma}{d E_\gamma} &=& 9.40 \times 10^{-12} \left( \frac{\langle \sigma v \rangle}{10^{-27} cm^3 \cdot s^{-1}}  \right)   \nonumber \\
&\times& \left( \frac{100 \; \mbox{GeV}}{m_\chi}  \right)^2 \sum_f \frac{dN_f}{dE_{\gamma}} B_f \; \langle J \rangle \Delta \Omega,
\eea
which is in units of photons$/$cm$^{2}$$/$s$/$GeV.

$\gamma$-ray signals from dark matter annihilations must compete with the diffuse $\gamma$-ray background.
These include inverse Compton scattering of electrons with galactic radiation and bremsstrahlung processes from accelerated charges \cite{Strong:2007nh}.  
Thus, uncertainties in the propagation of cosmic rays and in the composition of the ISM (e.g. the distribution and density of hydrogen) lead to uncertainties in the expected diffuse background.
It is vital to understand the diffuse background  in order to confirm (or deny) the existence of dark matter annihilations and to distinguish between different theoretical predictions. 

At the present level of understanding, the differential flux for the diffuse $\gamma$-ray background may be fitted by a power-law of the form \cite{Bergstrom:1997fj}
\be \label{bergback}
\frac{ d^2 \Phi_\gamma^{bg} }{d \Omega dE_{\gamma}  } = \left(  \frac{ d^2 \Phi_\gamma^{bg} }{d \Omega dE_{\gamma}  }  \right)_0  \left(  \frac{E_\gamma}{1 \mbox{GeV} } \right)^\alpha, 
\ee
with $\alpha = -2.72$  and a normalization $ \left(  { d^2 \Phi_\gamma^{bg} }/{d \Omega dE_{\gamma}  }  \right)_0 = 6 \times 10^{-5}$.   

\subsubsection{Constraints from EGRET}
EGRET completed nine years of observations in June of 2000 and was sensitive to $\gamma$-rays in the energy range $30$ MeV - $30$ GeV.
Using (\ref{bergback}) for the diffuse background near the galactic center and integrating over the angular resolution of EGRET  $(\Delta \Omega = 10^{-3})$  for the energy range of interest  (1 GeV $\lesssim E_\gamma \lesssim 30$ GeV) we find a background flux of $\Phi^{bg}_\gamma \simeq 10^{-8}$ photons cm$^{-2}$ s$^{-1}$.  For dark matter candidates that give a flux in excess of this, EGRET should have detected a signal.  From (\ref{diffflux}), a neutralino annihilating to $W$-bosons with a mass of a couple hundred GeV and cross section $\langle \sigma v \rangle \approx 10^{-24}$ cm$^{3}$ s$^{-1}$ yields a flux comparable to the background $\Phi^{dm}_\gamma \simeq 4 \times10^{-8}$ photons cm$^{-2}$ s$^{-1}$.   This gives the first indication of the tension between a 200 GeV wino and the EGRET data.  Of course, this result depends on the dark matter profile -- assumed here to be NFW.  

Extracting robust constraints on dark matter candidates from EGRET is subtle for reasons extending beyond the choice of the profile: there are uncertainties in both the diffuse background, as well as the EGRET data itself.

EGRET  has detected a possible excess above $1$ GeV in all sky directions. 
Addressing the discrepancy between the expected diffuse background and the EGRET data has been considered by a number of authors \cite{Dodelson:2007gd,Cesarini:2003nr,Fornengo:2004kj,Strong:2004de,Strong:2007nh,Stecker:2007xp,deBoer:2005bd,Bergstrom:2006tk}.  
These authors have argued for explanations that range from the possibility of annihilating dark matter\footnote{This explanation relies on a non-standard (anisotropic) choice for the halo profile, and seems to be at odds with other sources of indirect detection \cite{Bergstrom:2006tk}.}  \cite{deBoer:2005bd} to systematic errors in the EGRET experiment \cite{Stecker:2007xp}. 

A key challenge for addressing the possible excess is developing an accurate model of the astrophysical background.  This is particularly challenging given the inability to disentangle various components.  These include the weak extragalactic contribution to the diffuse background, as well as a number of possible unresolved point sources \cite{Dodelson:2007gd,Baltz:2008wd}.  Due to the uncertainties, proposed models for the background can vary significantly.  Compared to the background in Eqn.~(\ref{bergback}),

 the `conventional' GALPROP model \cite{Strong:1998pw} assumes a larger contribution from inverse Compton scattering, giving a higher contribution to the background and therefore to any signal that would be seen by EGRET.  Yet other choices of background exist, including the 'Optimized' background \cite{Strong:1998pw},  chosen to fit the EGRET excess without any additional Dark Matter component.  At present, the take-home message is that there are large uncertainties in the astrophysical background.

In addition to the uncertainties associated with the diffuse background and the halo profile, there are other reasons for concern in regards to the quality of the EGRET data.  Indeed, EGRET was only designed to operate for two of its nine year mission and an aging spark chamber introduced time-dependent uncertainties and systematic errors into the high end data products \cite{Moskalenko:2006zy}.  In \cite{Stecker:2007xp} is was found that the most likely explanation of the EGRET excess was an error in the estimation of the EGRET sensitivity at energies above a GeV.  This was argued to be convincing given that the `excess' is seen in all sky directions, not just towards the galactic center.  

With these caveats in mind, we use the EGRET data to constrain the 200 GeV wino.  We state the constraint as a bound on the $\langle J \rangle$. 
Assuming EGRET correctly measured the background above a GeV and using the data from \cite{MayerHasselwander:1998hg} to determine the diffuse background, we find that a $200$ GeV wino has an annihilation cross section too large by a factor of three for an NFW profile --  for a softer profile $\langle J \rangle \simeq 380$ this would not be the case.  These findings agree with already existing bounds in the literature \cite{Hooper:2002ru,Dodelson:2007gd,Baltz:2008wd}.  However, we have also found that using the lower choice for the diffuse background in Eqn.~(\ref{bergback})
implies that the $200$ GeV wino is already marginally consistent with the EGRET data for the NFW profile\footnote{See \cite{Altunkaynak:2008ry} for a similar approach to dealing with uncertainties in the diffuse background and bounds on neutralinos coming from EGRET and FGST.}.    

For now, it seems reasonable to consider the close proximity of the $200$ GeV dark matter to the current bounds set by EGRET encouraging, since we will see in Section \ref{section3} that the improvements of FGST should clarify the situation. 

\subsection{Anti-Proton Bounds}
\begin{figure}[!] 
\includegraphics[width=\columnwidth]{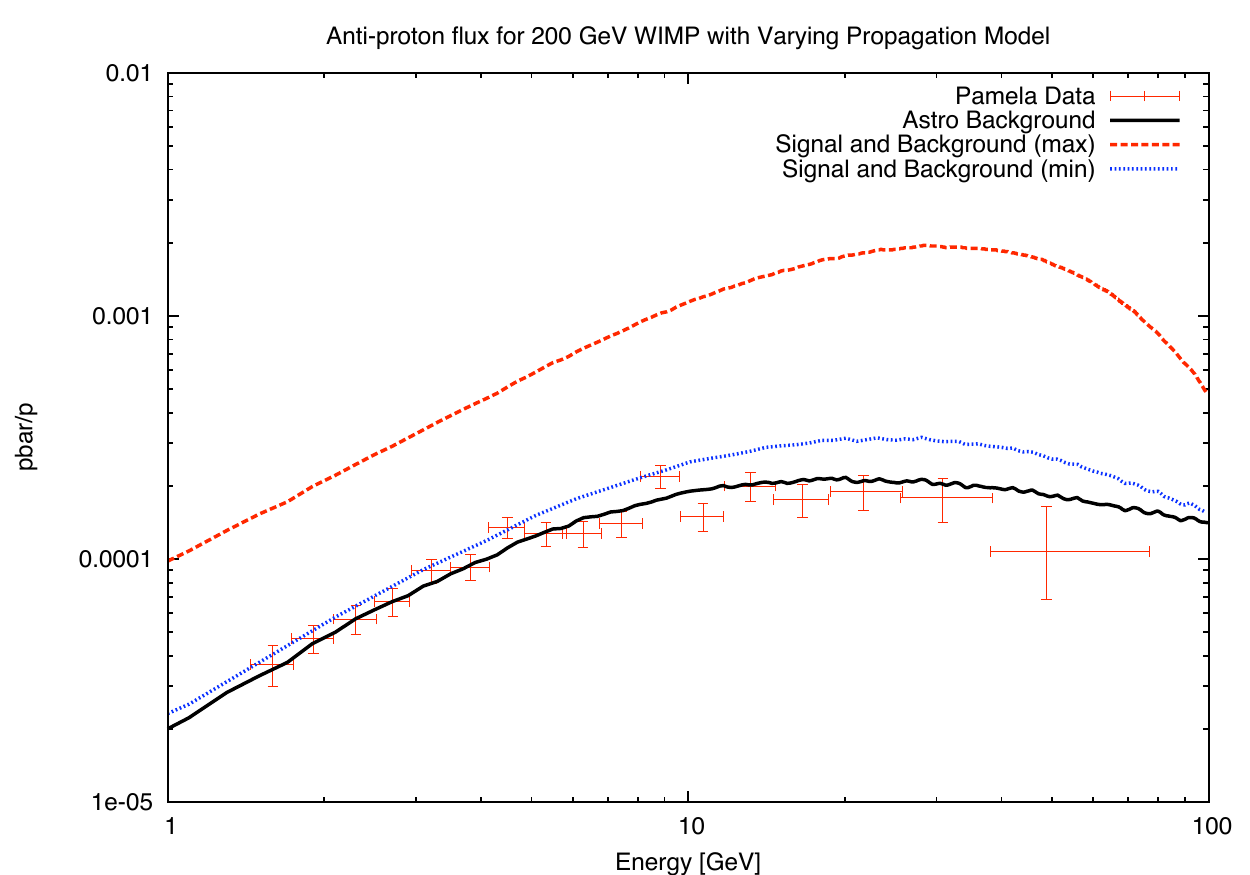}
\caption{\label{proton_bound} The anti-proton flux ratio for a $200$ GeV wino-like neutralino as a function of kinetic energy.  The lowest curve represents the conventional astrophysical background, whereas the remaining curves are the signal plus background for the $200$ GeV candidate.  These curves are the flux from dark matter annihilations given different choices for propagation model -- all of which have been parametrically fixed by matching to the well known spectrum of secondary/primary fluxes (e.g. B/C ratio) \cite{Donato:2003xg}.  }
\end{figure}

Measurements of cosmic ray anti-proton fluxes can also be used to put constraints on light neutralino candidates.
In fact, the PAMELA experiment will measure anti-proton fluxes in the energy range $80$ MeV - $190$ GeV.  It has already reported early data 
\cite{Boezio:2008mp} which seems consistent with and extends earlier results, e.g. \cite{BESS,Orito:1999re,Abe:2008sh}.

Taken at face value, the anti-proton data would appear to exclude a 200 GeV wino as an explanation of the PAMELA data, see e.g. \cite{Cirelli:2008pk}.     
However, anti-proton constraints suffer from theoretical uncertainties in cosmic ray propagation, as has been demonstrated in 
\cite{Donato:2003xg} (see also the discussion in \cite{Bergstrom:2006tk}).  One approach to bound the uncertainties and set propagation parameters for anti-protons is to parametrically fit models to well measured secondary/primary fluxes such as the Boron to Carbon (B/C) ratio.  In \cite{Donato:2003xg} it was found that this technique suffers from a number of degeneracies. 
These degeneracies arise from the fact that the effective size of the confining region in which the cosmic rays propagate and the amount of energy lost to diffuse processes can be varied in combination, giving a good fit to the B/C ratio for a variety of values.

\begin{figure}[!]  
\includegraphics[width=\columnwidth]{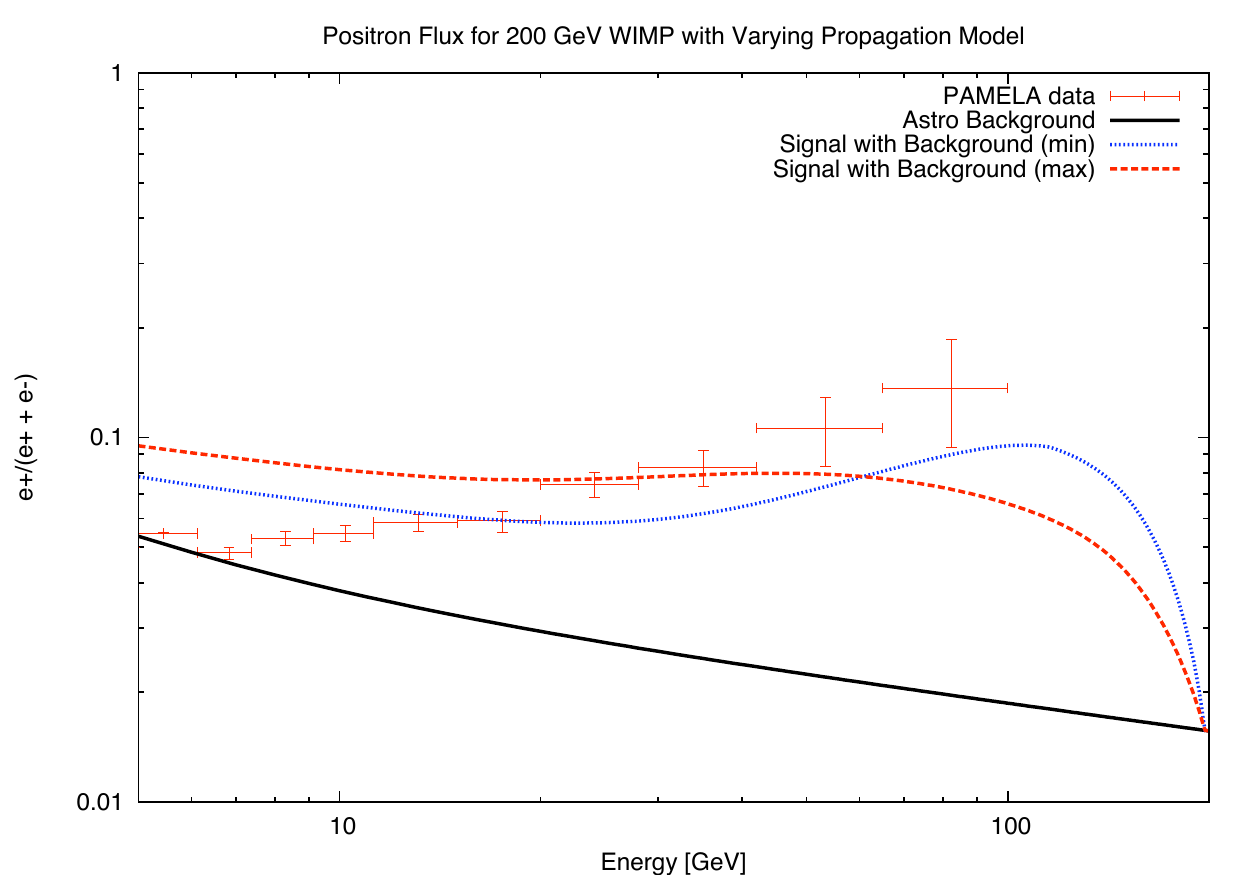}
\caption{\label{pos_prop2} Positron flux ratio for a wino-like neutralino with a mass of $200$ GeV.  The lowest curve represents the astrophysical background, whereas the remaining curves are the flux ratio for (large) energy loss rate of $\tau=5 \times 10^{16}$ s and varying propagation model (as discussed in the text). }
\end{figure}

For the $200$ GeV candidate we consider here, and assuming an NFW profile, the uncertainties in the propagation can lead to 
variations in the Dark Matter induced anti-proton flux by as much as an order of magnitude.  This can be seen in Figure \ref{proton_bound}, where we present the dark matter signal for the benchmark propagation models appearing in \cite{Donato:2003xg} that yield the minimum and maximum anti-proton signal.  Both models are consistent with the B/C ratio.  The order of magnitude variation in the theoretical prediction might cause the reader to be hesitant to conclude a 200 GeV wino is excluded from the data. At present, even for the minimal choice of propagation model, the $200$ GeV candidate still gives a prediction that is about two times that expected from the recent observations of PAMELA \cite{Boezio:2008mp}.  If a 200 GeV wino is to explain the data, there must be additional problems with the models used to propagate the anti-protons.

Variation of the propagation parameters will also influence the positron spectrum.  Once the anti-proton flux is minimized, what happens to the positrons? This effect is not that pronounced, primarily because the high-energy positrons relevant for PAMELA originate within a couple kpc of Earth.  Propagation uncertainties are thus reduced relative to those for anti-protons.  Fig.~\ref{pos_prop2} shows the effect on positrons of using the ``min" and ``max" models used for  Fig.~\ref{proton_bound}.

\section{Future Probes and Predictions for FGST \label{section3}}
As we have mentioned, PAMELA will probe positron cosmic rays from $50$ MeV up to an estimated  $270$ GeV.
Thus, if a light wino-like neutralino is responsible for the positron excess, PAMELA must see a turn-over 
in the data, as can be seen from Figure \ref{figurepos}.

The ATIC experiment \cite{:2008zz} has also reported an excess of cosmic ray electrons above roughly $300$ GeV.  The wino candidate we describe here could not be responsible for this excess.    If the $200$ GeV wino indeed accounts for the PAMELA excess, another explanation would be required for the ATIC data. 

We now consider the ability of FGST to detect the 200 GeV wino invoked above 
 We focus on measurements of the galactic center, though measurements of the halo could be useful if progress is made in understanding the backgrounds there in detail.

FGST will offer a significant improvement over EGRET, probing energies from $20$ MeV to $300$ GeV with an angular resolution of around $0.1$ degrees ($\approx 10^{-5}$ sr).
The improved angular resolution will not only allow for separation of the point sources detected by EGRET, but the increased sensitivity will allow for a better opportunity to distinguish dark matter annihilation signals from the diffuse background.  For the energy range of interest (around 1 GeV $\lesssim E_\gamma \lesssim 300$ GeV) one finds a background flux from (\ref{bergback}) of around $\Phi^{bg}_\gamma \simeq 10^{-10}$ photons cm$^{-2}$ s$^{-1}$ at a FGST angular acceptance of $10^{-5}$ sr.  Compared with the EGRET result of $\Phi^{EGRET}_\gamma \simeq 10^{-8}$ photons cm$^{-2}$ s$^{-1}$, this allows for an improved sensitivity by two orders of magnitude in terms of resolving signal from background.  As in the case of EGRET, the diffuse background and halo profile are both sources of significant uncertainty. The better resolution and ability of FGST to resolve point sources should improve our knowledge of the diffuse background. 

\begin{figure}[t!]
\includegraphics[width=\columnwidth]{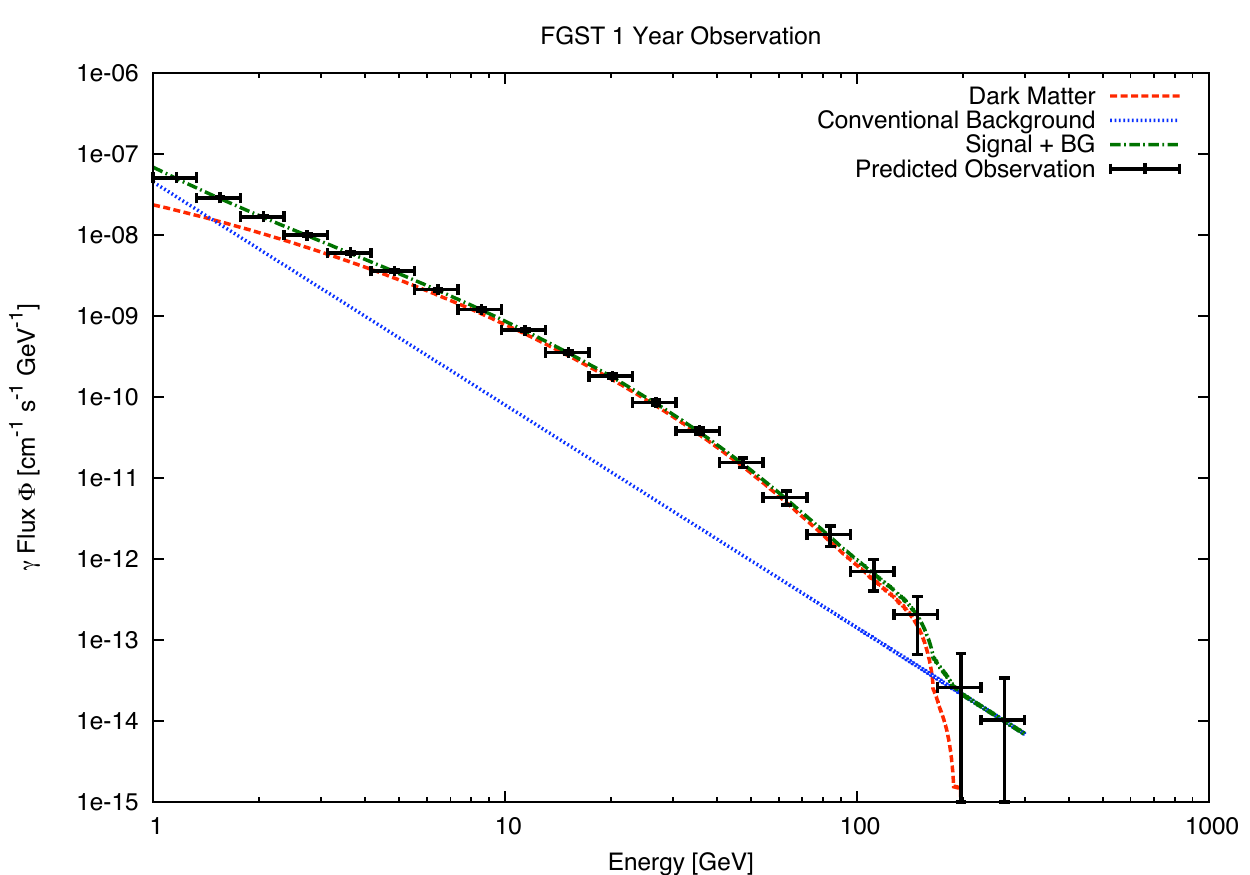}
\caption{The differential flux for the $200$ GeV wino-like neutralino and an NFW profile averaged over the minimum angular resolution of FGST (i.e. $\Delta \Omega = 10^{-5}$ sr) and integrated over a $0.5^{\circ} \times 0.5^{\circ}$ region around the galactic center. For the diffuse background we take the `conventional' galprop model discussed in the text.  The error bars represent the statistical uncertainty after one year of observations and do not account for systematical uncertainties. \label{glast200}}
\end{figure}

For our predictions for FGST we consider an $0.5^\circ \times 0.5^\circ$ region about the galactic center assuming an NFW profile and averaging with a minimum resolution set by FGST (i.e. $10^{-5}$ sr).  We have considered a number of choices for modeling the diffuse background.  We find that for both a low choice of background given by the power-law with normalization in (\ref{bergback}), as well as for higher backgrounds such as the `conventional' and `optimized' backgrounds mentioned above, that FGST will report a signal early.  We find for the conventional background and a $200$ GeV wino that a variation in the halo profile down to $\langle J \rangle \simeq 70$ in the region about the galactic center can still result in a detectable signal for FGST at the $5\sigma$ level within the first year of observation.  In Figure \ref{glast200} we present the prediction for the $200$ GeV wino with an NFW profile, again after only one year of data.  The error bars reflect statistical uncertainties.  FGST should be capable of observing the products of wino annihilation after the first year.  

\section{Conclusions and Outlook}
The PAMELA group \cite{Adriani:2008zr} has reported a robust excess in galactic positrons with energy above about 10 GeV, compared to those expected from known astrophysical sources.  
This is consistent with earlier reported excesses from \cite{Barwick:1995gv,Barwick:1997ig,AMS01}. The source that produces postirons will likely produce antiprotons and $\gamma$-rays.  No unambiguous excess in these channels has been observed, which provides constraints on a dark matter interpretation. 

 Only if standard assumptions about the propagation of cosmic rays are relaxed can the excess in positrons arise from galactic annihilation of winos, the superpartners of W bosons.  
Although there are apparent conflicts with anti-protons and gammas, given the astrophysical uncertainties it is likely premature to assume the wino interpretation is completely excluded.  Further investigation of these uncertainties is a topic for future work.  Fortunately, data from FGST and further  data from PAMELA will help to clarify the picture soon.  Later data from AMS-02 will also help.
 
If the PAMELA excess is indeed due to a well-motivated wino LSP, the implications are remarkable.   We would learn what the dark matter of our universe is.  It would be the discovery of supersymmetry, telling us something about the resolution to the hierarchy problem.  It would imply a number of superpartners will likely be seen at LHC, confirming the result. 
And we would be learning that the universe had a non-thermal cosmological history  that we can probe.

\vspace{-.2in}
\section*{Acknowledgments}
\vspace{-.2in}
We thank Mirko Boezio, John Beacom, Daniel Cumberbatch, Jennifer Siegal-Gaskins, Igor Moskalenko, Brent Nelson, Malcolm Perry, Piergiorgio Picozza, Troy Porter, Pearl Sandick, Glenn Starkman, Greg Tarle, Todd Thompson, and Anna Zytkov for useful conversations. 
SW would like to thank Case Western and CCAPP for hospitality while this work was being completed.

This work is supported  by the Michigan Center for Theoretical Physics and the DOE under grant DE-FG02-95ER40899.The work of AP is also supported by NSF CAREER grant NSF-PHY-0743315.  The work of SW is also supported by the Michigan Society of Fellows.


\end{document}